\documentclass[apjl]{emulateapj}

\usepackage{natbib}
\usepackage{amsmath}
\usepackage{amssymb}
\usepackage[bf, sf, FIGTOPCAP, nooneline, tight]{subfigure}
\usepackage{graphicx}
\usepackage{dcolumn}
\usepackage{bm}
\usepackage[usenames,dvipsnames]{color}
\usepackage[colorlinks=true, linkcolor=BrickRed, citecolor=Blue, urlcolor=Blue, filecolor=Blue]{hyperref}
\usepackage{booktabs}
\usepackage{hhline}

\def\beq{\begin{equation}}
\def\eeq{\end{equation}}
\def\ber{\begin{eqnarray}}
\def\eer{\end{eqnarray}}

\def \lleq {\lower0.9ex\hbox{ $\buildrel < \over \sim$} ~}
\def \ggeq {\lower0.9ex\hbox{ $\buildrel > \over \sim$} ~}

\def\deg{\ifmmode^\circ\else$^\circ$\fi}

\newcommand{\apar}{\alpha_\parallel}
\newcommand{\aperp}{\alpha_\perp}
\begin{document}

\title{Test $\Lambda$CDM model with High Redshift data from Baryon Acoustic Oscillations}

\author{Yazhou Hu\altaffilmark{1,2}, Miao Li\altaffilmark{3,1}, Zhenhui Zhang\altaffilmark{1,2}}

\altaffiltext{1}{State Key Laboratory of Theoretical Physics, Institute of Theoretical Physics, Chinese Academy of Sciences, Beijing, 100190}
\altaffiltext{2}{Kavli Institute for Theoretical Physics China, Chinese Academy of Sciences, Beijing, 100190}
\altaffiltext{3}{School of Astronomy and Space Science, Sun Yat-Sen University, Guangzhou 510275, People's Republic of China}

\email{asiahu@itp.ac.cn}
\email{mli@itp.ac.cn}
\email{zhangzhh@itp.ac.cn}

\begin{abstract}

The Baryon Acoustic Oscillations (BAO) provide a standard ruler for studying cosmic expansion. The recent observations of BAO in SDSS DR9 and DR11 take measurements of $H(z)$ at several different redshifts. It is argued that the behavior of dark energy could be constrained more effectively by adding high-redshift Hubble parameter data, such as the SDSS DR11 measurement of $H(z) = 222\pm7$ km/sec/Mpc at z = 2.34.
In this paper, we investigate the significance of these BAO data in the flat $\Lambda$CDM model, by  combining them with  the recent observational data of the Hubble constant from local distance ladder and the Cosmic Microwave Background (CMB) measurements from Planck+WP. We perform a
detailed data analysis on these datasets and find that the recent observations of BAO in SDSS
DR9 and DR11 have considerable tension with the Planck + WP measurements in
the framework of the standard $\Lambda$CDM model. The fitting results show that the
main contribution to the tension comes from the Hubble parameter measurement at redshift of $z=2.34$. But there is no visible tension once the joint data analysis  by combining the datasets of  SDSS and Planck+WP is performed. Thus in order to see whether dark energy does evolve,
we need more independent measurements of the Hubble parameter at high redshifts.

\end{abstract}


\maketitle

\section{Introduction}

The cosmic acceleration was discovered in 1998 \citep{Riess1998,Perlmutter1999}, and it has  been confirmed by several independent observations.
The standard $\Lambda$CDM model provides a succinct description of the cosmic acceleration matching current
Cosmic Microwave Background (CMB) observations \citep{planck2013}.
Besides the CMB  measurements, the Baryon Acoustic Oscillation (BAO) measurements are recognized as robust and independent probes of cosmology, since they provide a standard ruler for studying the cosmic expansion.
Recently,  \citet{bao2014} reported a detection of the BAO feature in the flux-correlation function of the Ly$\alpha$ of high-redshift quasars from the Data Release 11 (DR11) of the Baryon Oscillation Spectroscopic Survey
(BOSS) \citep{Dawson2013} of SDSS-III \citep{Eisenstein2011}.  By adopting the value of sound horizon at the drag epoch $r_d=147.4$ Mpc from the Planck+WP fitting of the concordance cosmology, 
\citet{bao2014} derive a high-redshift Hubble parameter $H(z=2.34)=222\pm7$ km/sec/Mpc.
As the measurements of  $H(z)$ parameter at different redshifts are statistically independent,
they are suitable for studying the evolution of the cosmic geometry.
With these high-redshift $H(z)$ data,  \citet{sahni2014} test the cosmological constant hypothesis by employing the evolution of $H(z)$ according to the recent observations of BAO's in SDSS DR9 \citep{bao2013} and DR11 \citep{bao2014}.
By adopting an improved version of the $Om$ diagnostic,  \citet{sahni2014} find that there is considerable tension with the Planck+WP measurements in the framework of the standard $\Lambda$CDM model.
By the $Om$ diagnostic, the $Omh^2$ value is independent of the redshift if dark energy is a constant.
However, they find that $Om(z=2.34)h^2=0.122\pm0.01$, while the Planck result is $Omh^2=0.1426\pm 0.0025$ \citep{planck2013}.
Therefore, the $\Lambda$CDM model is in conflict with the Hubble parameter measurement at redshift $z=2.34$, at the level of two standard deviation.

In this paper, we investigate this tension through joint data analysis instead of the $Omh^2$ diagnostic, in order to compare the significance of these high-redshift BAO data with that of the Planck data.

\section{Methodology \& Results}

For comparison, we use the same $H(z)$ dataset of \citep{sahni2014}, including the Hubble constant from local distance ladder ($H(z=0) = 70.6 \pm 3.2$ km/sec/Mpc \citep{efstathiou_2014}), the Hubble parameter from anisotropic clustering of SDSS DR9 ($H(z=0.57) = 92.4 \pm 4.5$ km/sec/Mpc \citep{Reid2012}) and the new $H(z)$ data from the $Ly\alpha$ forest of SDSS DR11 quasars ($H(z)=222\pm7$ km/sec/Mpc at $z=2.34$ \citep{bao2014}).

As we noted above, the high-redshift Hubble parameter $H(z=2.34)$ is derived by scaling at a $r_d$ from the Planck+WP measurements,
 so the value of $H(z=2.34)$ is correlated with the Planck+WP dataset.
 To avoid this correlation, we also adopt the most precisely determined combination from
\citep{bao2014}, namely

\begin{equation}
\apar^{0.7}\aperp^{0.3} = 1.025 \pm 0.021 \;,
\end{equation}

where $\apar$ and $\aperp$ are defined as
\begin{equation}
\apar = \frac { \left[D_H(\bar z)/r_d\right] }{\left[D_H(\bar z)/r_d\right]_{\rm fid}}
\hspace*{3mm}{\rm and}\hspace*{5mm}
\aperp = \frac { \left[D_A(\bar z)/r_d\right] }{\left[D_A(\bar z)/r_d\right]_{\rm fid}} ~,
\label{eq:alpha}
\end{equation}
where the fiducial values $\left[D_H(\bar z)/r_d\right]_{\rm fid}$ and $\left[D_A(\bar z)/r_d\right]_{\rm fid}$ at $\bar z=2.34$ are 8.708 and 11.59 respectively. For the SDSS DR9 dataset, the result that $D_V(0.57)/r_s(z_d)=13.67\pm0.22$ obtained by \citet{Anderson2012} will be used. For a further comparison, we also employ the improved $D_V/r_s(z_d)$ measurements of SDSS DR11 dataset, which gives $D_V(z=0.32)/r_s(z_d)=8.25\pm0.17$ and $D_V(z=0.57)/r_s(z_d)=13.42\pm0.13$ \citep{Anderson2013} respectively.

In the following context,
we will use ``H0'', ``Hz'', ``DR9'', ``DR11'',``DR11Ly$\alpha$'', and ``Planck+WP'' to represent the Hubble constant, the three $H(z)$ data,  $D_V(0.57)/r_s(z_d)$  of SDSS DR9, $D_V(0.32)/r_s(z_d)$ and $D_V(0.57)/r_s(z_d)$  of SDSS DR11,  $\apar^{0.7}\aperp^{0.3}$  of SDSS DR11 at reshift $z=2.34$ and the Planck+WP dataset respectively.
For convenience, we will use ``SDSS'' to represent the combination of ``Hz'', ``DR9'', ``DR11'' and ``DR11Lya'' datasets.

With these datasets, we perform $\chi^2$ analysis and explore the parameter space using Markov Chain Monte Carlo (MCMC) algorithm by modifying the CosmoMC package~\citep{cosmomc}.

\begin{figure}[!t]
\centering{
\includegraphics[width=7.5cm,height=6.8cm]{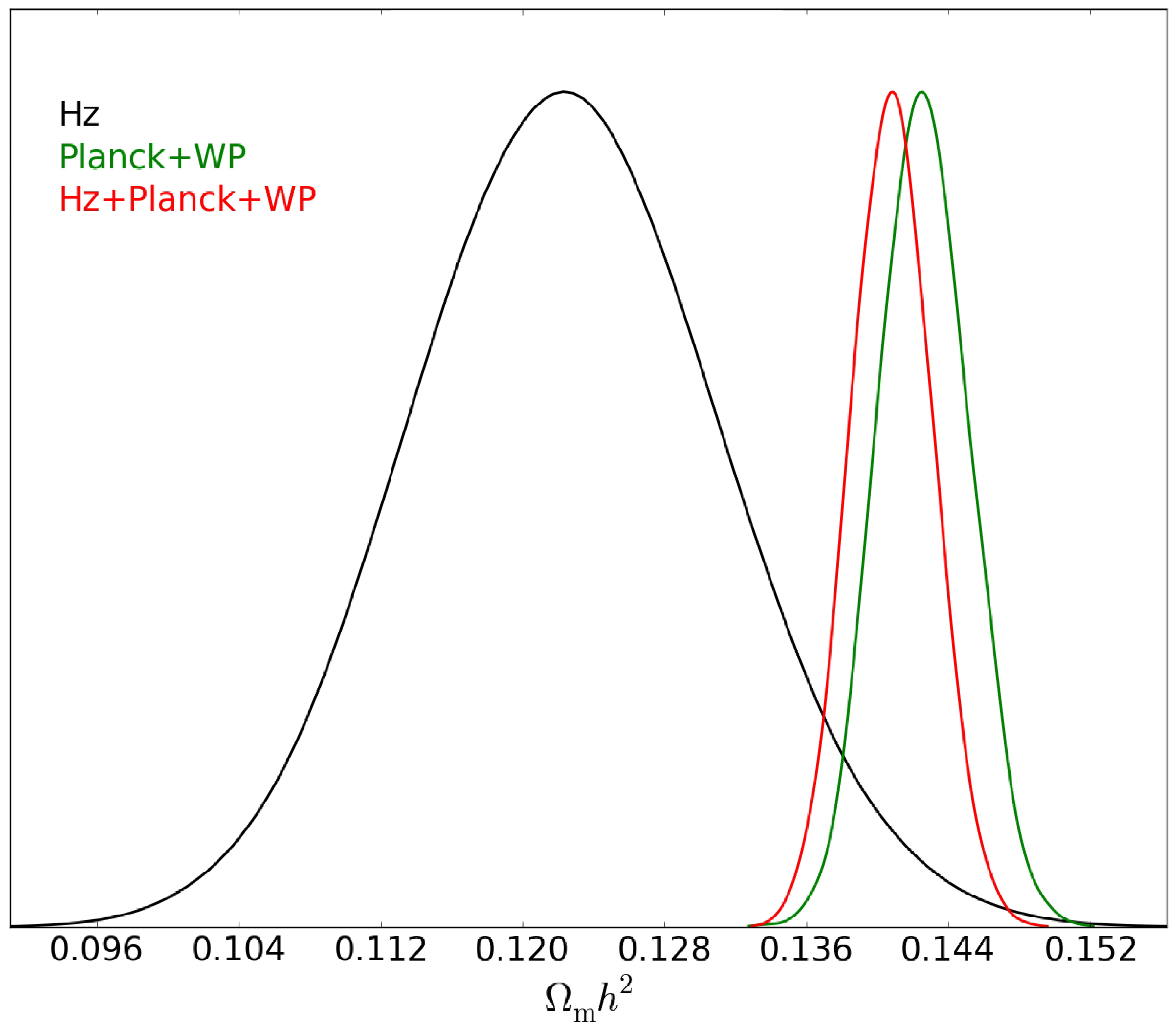}
\includegraphics[width=7.5cm,height=6.8cm]{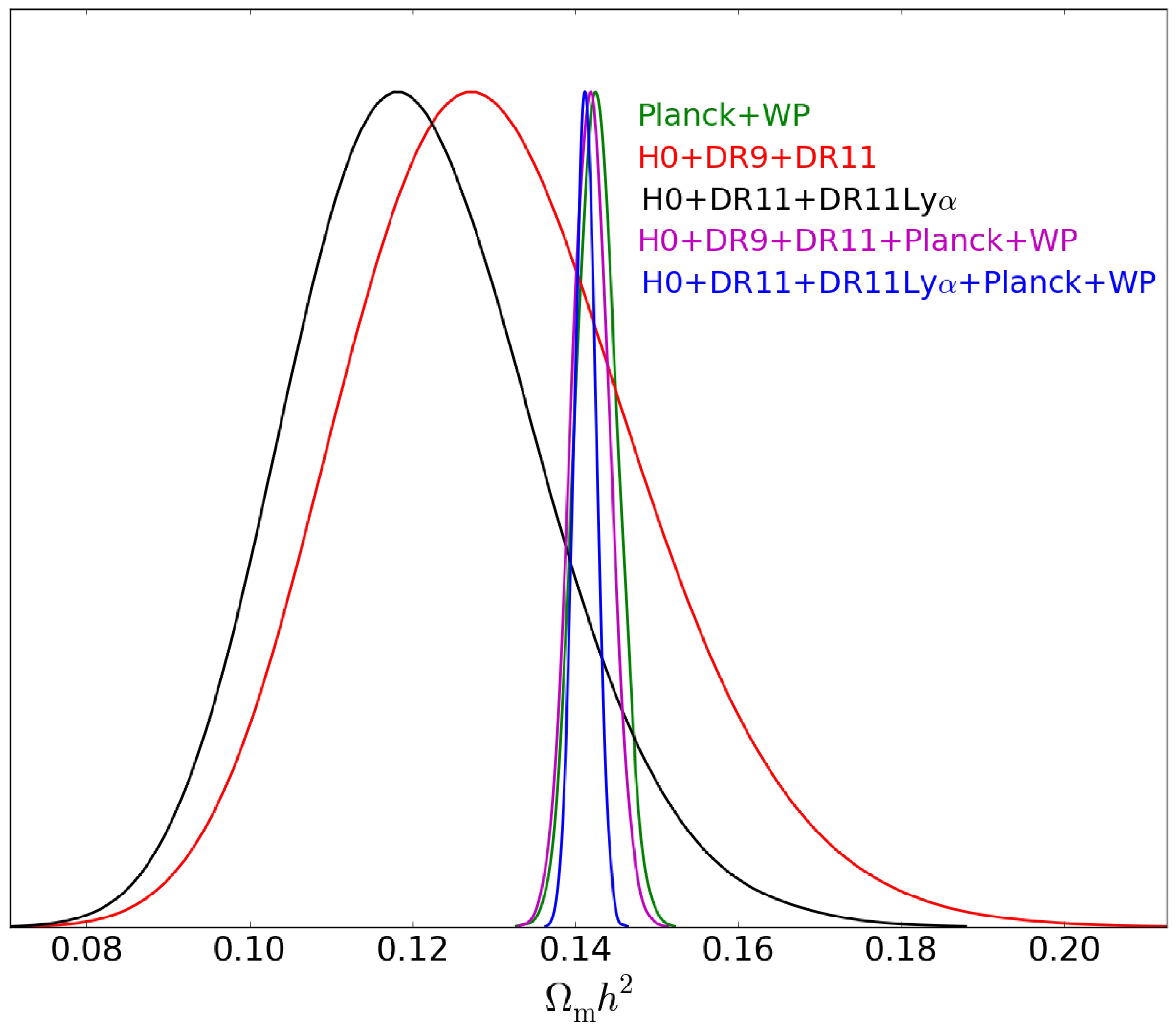}}
\caption{\label{fig1} (color online). Marginalized likelihood distributions of $\Omega_m h^2$ with different datasets as masked: in the upper panel, with the Hz and Planck+WP data and their combination; in the lower panel, with the  DR9, DR11 and Planck+WP data and their combinations }
\end{figure}

Table \ref{Table1} and Table \ref{Table2} summarize the fitting results of the parameter constraints (the 68\% CL limits) and corresponding $\chi^2_{min}$'s.  In Figure \ref{fig1}, in order to make a comparison, we plot the likelihood distributions  of $\Omega_m h^2$ with two different selections : the Hz and Planck+WP datasets (corresponding the upper panel) ; the  DR9, DR11 and Planck+WP datasets (corresponding the  below panel).

\begin{table*} \caption{Fitting results of $\Lambda$CDM model with the Hubble parameters datasets .}
\begin{center}
\label{Table1}
\begin{tabular}{cccccc}
  \hline\hline
  Dataset   &           Hz    &  $Planck$+WP  &    H0+H(z=0.57)+$Planck$+WP &  Hz+$Planck$+WP     \\
  \hline
  $\Omega_{m}h^{2}$ &   $0.1221\pm 0.0085^a $&  $0.1426\pm0.0025 $ &  $0.1419\pm0.0023 $  &  $ 0.1408\pm0.0023 $  \\
   $\chi^2_{min}$   &  $0.003$~~ &  $9804.166$~~&    $9806.278 $~~  & $9810.524 $              \\

  \hline\hline
\end{tabular}
\end{center}
\leftline{\noindent$\ ^a$ We list the 68\% CL limits.}
%\leftline{\noindent$\ ^b$ Corresponding fittings in $w$CDM model.}
\end{table*}

\begin{table*} \caption{Fitting results of the $\Lambda$CDM model with the SDSS datasets.}
\begin{center}
\label{Table2}
\begin{tabular}{cccccc}
  \hline\hline
  Dataset   &    H0+DR9+DR11Ly$\alpha$    &    H0+DR9+$Planck$+WP &  H0+DR9+DR11Ly$\alpha$+$Planck$+WP  & H0+DR11+DR11Ly$\alpha$+$Planck$+WP  \\
  \hline
  $\Omega_{m}h^{2}$ &   $0.1305\pm 0.0179^a  $ &  $0.1423\pm0.0019 $  &  $ 0.1419\pm0.0023 $ & $0.1413\pm0.0013$ \\
   $\chi^2_{min}$   &  $0.0002$~~&    $9806.092 $~~  & $9808.850 $ & $ 9811.142$            \\

  \hline\hline
\end{tabular}
\end{center}
\leftline{\noindent$\ ^a$ We list the 68\% CL limits.}
\end{table*}

\citet{sahni2014} show that the estimation of the new diagnostic $Omh^2$ from SDSS DR9 and DR11 data gives $\Omega_{m}h^{2} \approx 0.122 \pm 0.01$, having tension with the value
$\Omega_{0m}h^2 = 0.1426 \pm 0.0025$ determined for $\Lambda$CDM from Planck+WP at over $2 \sigma$. From Table \ref{Table1} and the upper panel of Figure \ref{fig1}, we can see that the Hz and the Planck+WP give constraints
$\Omega_{m}h^2 = 0.1221\pm 0.0085$ and $\Omega_{m}h^2 = 0.1426\pm 0.0025$ respectively, which means the Hz dataset does have tension with the Planck+WP dataset at over $2 \sigma$.
However, when using the Hz+Planck+WP and H0+H(z=0.57)+Planck+WP datasets, the constraints are $0.1408\pm0.0023$ and $0.1419\pm0.0023$  respectively. Apparently, they are both consistent with the constraint of Planck+WP dataset.

  Table \ref{Table2}  and the lower panel of  Figure \ref{fig1} show the constraints with the corresponding DR9, DR11, DR11Ly$\alpha$ and Planck+WP datasets.   When we use the $D_V(0.57)/r_s(z_d)$ measurement of SDSS DR9 in place of the Hubble parameter $H(z=0.57)$, the tension is alleviated, this is because the measurement of $H(z=0.57)$ of \citep{Reid2012} includes the anisotropic information of galaxy clustering while DR9 measurement of \citep{Anderson2012} does not. Thus the constraint on $\Omega_m h^2$ of $H(z=0.57)$ is tighter than DR9 measurement.
  For a further comparison,  we replace the DR9 data $D_V(0.57)/r_s(z_d)=13.67\pm0.22$ with the improved measurements $D_V(0.32)/r_s(z_d)=8.25\pm0.17$ and $D_V(0.57)/r_s(z_d)=13.42\pm0.13$ of DR11 data, the constraint becomes a little tighter, but the tension is still alleviated when compared with the constraint by $H(z=0.57)$ data.

An obvious feature can be seen from both Table \ref{Table1} and Table \ref{Table2},  that is the contribution to $\chi^2_{min}$'s is mainly dominated by the Planck+WP data.  These $\chi^2_{min}$'s just change relatively small values by adding these SDSS data. The results reveal that the high-redshift BAO data
is of less importance compared with the Planck+WP data on constraining the standard $\Lambda$CDM model.

%\begin{figure*}[!t]
%\centering
%\begin{center}
%\vspace{-0.00in}
%\centerline{\mbox{\hspace{0.in} \hspace{1.4in}  \hspace{1.4in} }}
%$\begin{array}{@{\hspace{0.0in}}c@{\hspace{0.5in}}c@{\hspace{0.5in}}c}
%\multicolumn{1}{l}{\mbox{}} &
%\multicolumn{1}{l}{\mbox{}} \\ [0.5cm]
%\includegraphics[scale=0.5, angle=0]{omegamh2.eps}
%\end{array}$
%\end{center}
%\caption{(color online). Marginalized likelihood distributions of the $\Omega_m h^2$ with different datasets as masked. }
%\label{fig2}
%\end{figure*}%

\begin{figure*}[!t]
\centering
\begin{center}
\vspace{-0.00in}
\centerline{\mbox{\hspace{0.in} \hspace{1.4in}  \hspace{1.4in} }}
$\begin{array}{@{\hspace{0.0in}}c@{\hspace{0.5in}}c@{\hspace{0.5in}}c}
\multicolumn{1}{l}{\mbox{}} &
\multicolumn{1}{l}{\mbox{}} \\ [0.5cm]
\includegraphics[scale=0.48, angle=0]{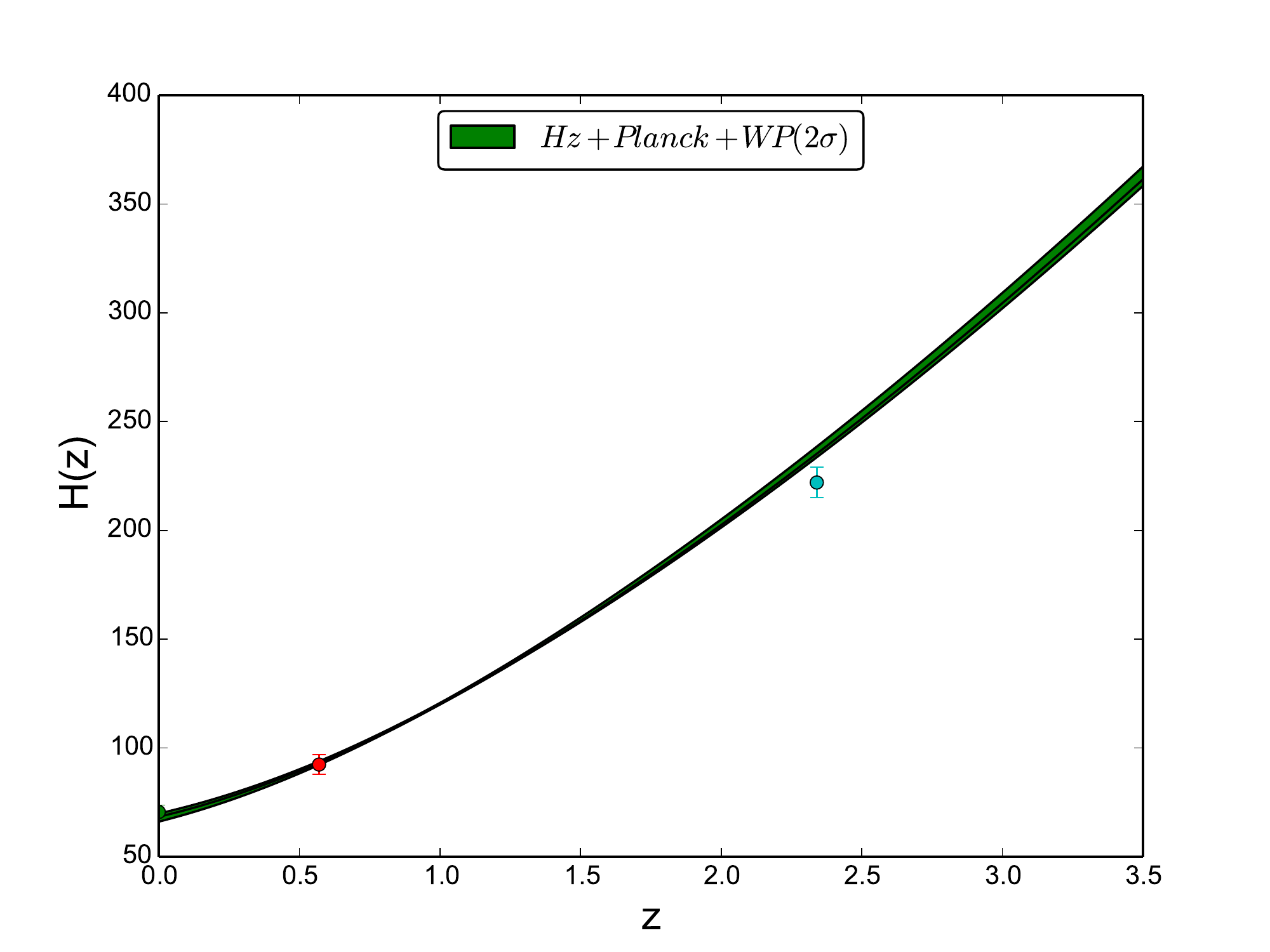}
\end{array}$
\end{center}
\caption{The evolution of Hubble parameter in the $\Lambda$CDM model constrained by the SDSS+Planck+WP data.
The best-fit (black line) and  2$\sigma$ constraints (filled region) are shown. Three $H(z)$ data points with their error bars are marked as well. }
\label{fig2}
\end{figure*}

For the $\Lambda$CDM model,the best-fit (black line) and 2$\sigma$ constraints (filled region) of the reconstructed evolution of $H(z)$ constrained by the Hz+Planck+WP data are plotted in Figure~\ref{fig2} respectively.
The three $H(z)$ data points with their error bars are marked as well.
We find that the $H(z)$ value given by high redshit BAO data deviates from the reconstructed 2$\sigma$ region of $H(z)$ constrained by the Hz+Planck+WP dataset.

We notice that there are similar features  described in \citep{sahni2014}. This is consistent with the results shown in Figure~\ref{fig1}: the difference between the constraints by Planck+WP data and that by combining with other SDSS datasets is not much. So the constraints of $\Omega_m h^2$ are also mainly dominated by the Planck+WP data. We conclude that the  significance of these high-redshift BAO data is less than  that of the Planck data.

\section{Summary \& Conclusion}

Using only the SDSS data, the contribution to the Hubble parameter constraint is mainly dominated by the $H(z=2.34)$ data. However, when we combine the SDSS and the Planck results,
the fitting is found to be mostly influenced by the low redshift $H(z)$ data and the Planck data, so the $\Lambda$CDM model looks fine. These results indicate that the single high redshift measurement
of $H(z)$ is at variance with other data, in order to see whether dark energy does evolve,
we need more independent measurements of the Hubble parameter at hight redshifts.

\begin{acknowledgments}
ML is supported by the National Natural Science Foundation of China (Grant No. 11275247, and Grant No. 11335012) and 985 grant at Sun Yat-Sen University. YH and ZZ would like to thank the support of the School of Astronomy and Space Science, Sun Yat-Sen University.
\end{acknowledgments}


\begin{thebibliography}{}
\expandafter\ifx\csname natexlab\endcsname\relax\def\natexlab#1{#1}\fi

\bibitem[Planck XVI (2013)]{planck2013}
Ade, P. {et~al.}, 2014, Astronomy and Astrophysics 571, A16

\bibitem[Anderson {et~al.} (2012)]{Anderson2012}
Anderson, L. {et al.}, 2012, MNRAS 427, 3435A

\bibitem[Anderson {et~al.} (2013)]{Anderson2013}
Anderson, L. {et al.}, 2013, arXiv:1312.4877

\bibitem[Dawson {et al.}(2013)]{Dawson2013}
Dawson, K. {et al.}, 2013,  Astron. J., 145, 10

\bibitem[Delubac {et~al.} (2014)]{bao2014}
Delubac, T. {et al.}, 2014, arXiv:1404.1801

\bibitem[Efstathiou (2014)]{efstathiou_2014}
Efstathiou, G. 2014, MNRAS 440, 1138

\bibitem[Eisenstein {et~al.} (2011)]{Eisenstein2011}
Eisenstein, D.J. {et al.}, 2011, Astron. J., 142, 721

\bibitem[Lewis \& Bridle (2002)]{cosmomc}
Lewis, A., \& Bridle, S., 2002, Phys.\ Rev.\ D, 66, 103511

\bibitem[Perlmutter {et al.}(1999)]{Perlmutter1999}
Perlmutter S. {et al.},  Astrophys. J. 517 (1999) 565

\bibitem[Reid {et al.} (2012)]{Reid2012}
Reid B. A. {et al.} MNRAS 426,  2719

\bibitem[Riess {et al.} (1998)]{Riess1998}
Riess A. G. {et al.}, Astron. J. 116 (1998) 1009

\bibitem[Sahni  {et~al.} (2014)]{sahni2014}
Sahni,V. {et al.}, 2014 , Astrophys.J. 793 (2014) L40

\bibitem[Samushia {et~al.} (2013)]{bao2013}
Samushia, L. {et al.}, 2013, MNRAS 429, 1514






\end{thebibliography}
\end{document}